\documentclass[twocolumn,superscriptaddress,showpacs,nofootinbib,notitlepage,secnumarabic,amssymb, nobibnotes,aps,prd,preprintnumbers]{revtex4-2}
\usepackage[utf8]{inputenc}
\usepackage{ulem}
\usepackage{graphicx}
\usepackage{latexsym,amsmath,amssymb,lmodern,float,url}
\usepackage{natbib}
\usepackage{color}
\usepackage{microtype}
\usepackage{import}
\usepackage{bbold}
\usepackage[plain]{fancyref}
\usepackage{varioref}
\usepackage{slashed}
\usepackage{multirow}
\usepackage{tikz}
\usetikzlibrary{shapes}

\usepackage[colorlinks=true,backref=false, linktocpage=true,
citecolor=blue,urlcolor=blue,linkcolor=blue,pdfpagemode=UseOutlines]{hyperref}

\hypersetup{%
  bookmarksnumbered=true,
  pdftitle = {},
  pdfsubject = {},
  pdfauthor = {},
  pdfkeywords = {}
}

\DeclareMathOperator{\Tr}{Tr}

\DeclareMathOperator{\re}{Re}

\definecolor{mycolor}{RGB}{22,139,22}

\newcommand{\bluetri}{\raisebox{0.7pt}{\tikz{\node[fill,scale=0.4,regular polygon, regular polygon sides=3,fill=blue!10!blue,rotate=0](){};}}}
\newcommand{\blacktri}{\raisebox{0.5pt}{\tikz{\node[fill,scale=0.4,regular polygon, regular polygon sides=3,rotate=180](){};}}}
\newcommand{\rbcir}{\raisebox{0.5pt}{\tikz{\node[draw,scale=0.5,circle,draw=red](){};}}}
\newcommand{\redfcir}{\raisebox{0.7pt}{\tikz{\node[draw,scale=0.5,circle,draw=red,fill=red,rotate=0](){};}}}
\newcommand{\bluesq}{\raisebox{0.7pt}{\tikz{\node[draw,scale=0.5,regular polygon, regular polygon sides=4,draw=blue,rotate=0](){};}}}

\begin{document}
\preprint{FERMILAB-PUB-20-611-T}
\title{Toward Quantum Simulations of $\mathbb{Z}_2$ Gauge Theory Without State Preparation}
\author{Erik  J.  Gustafson}
\email{erik-j-gustafson@uiowa.edu}
\affiliation{Department of Physics and Astronomy, The University of Iowa, Iowa City, IA 52242, USA}
\author{Henry Lamm}
\email{hlamm@fnal.gov}
\affiliation{Fermi National Accelerator Laboratory, Batavia,  Illinois, 60510, USA}
\begin{abstract}
Preparing strongly-coupled particle states on quantum computers requires large resources.  In this work, we show how classical sampling coupled with projection operators can be used to compute Minkowski matrix elements without explicitly preparing these states on the quantum computer.  We demonstrate this for the 2+1d $\mathbb{Z}_2$ lattice gauge theory on small lattices with a quantum simulator. 
\end{abstract}

\maketitle


\section{Introduction}
\label{sec:intro}

Quantum computers hold the promise of solving problems beyond the capability of classical computers in many aspects of high energy physics, in particular calculations involving finite density and real time evolution~\cite{Feynman:1981tf}. Achieving this promise requires the development not only of hardware with reduced noise but efficient algorithms  as well.  At present, a major roadblock to such simulations is the creation of the initial, strongly-coupled quantum states on the quantum computer. Current methods for this state preparation are expensive to implement (usually dominating the circuit depth of simulations) and often difficult to generalize beyond ground states. In particular, the preparation of scattering states requires substantial complexity~\cite{Jordan:2011ne,Jordan:2011ci,Garcia-Alvarez:2014uda,Jordan:2014tma,Jordan:2017lea,Moosavian:2017tkv,Moosavian:2019rxg,Gustafson:2019mpk,Gustafson:2019vsd}. Consequently, attempts to develop more efficient procedures have been explored~\cite{Kokail:2018eiw,Lamm:2018siq,Klco:2019xro,Klco:2019yrb,Harmalkar:2020mpd}. 

One proposal~\cite{Lamm:2018siq,Harmalkar:2020mpd} avoids the issue of initial state preparation entirely by  stochastically sampling the density matrix, $\rho$, classically, and then passing simpler basis states to the quantum computer with proper weights. This method can be understood as a Schwinger-Keldysh contour~\cite{Schwinger:1960qe,Keldysh:1964ud}, where the classically sampled Euclidean path integral is matched at its boundaries to a Minkowski path integral computed on the quantum computer.  For quantum field theories, the natural way to sample $\rho$ is through a lattice field theory (LFT) calculation with open boundary conditions. In this picture, the quantum computer acts as an operator insertion into a standard LFT calculation, except it returns time-dependent quantities. For thermal states, one need only sample from the Euclidean path integral at the desired inverse temperature $\beta$~\cite{Lamm:2018siq}. For other states (e.g. the pion or scattering protons), it was proposed to use projections of the configurations onto the quantum numbers of the desired states~\cite{Harmalkar:2020mpd}. 

In this paper, we demonstrate the nonthermal state preparation using projection operators suggested in~\cite{Harmalkar:2020mpd} by computing results for one- and two-``particle" plane waves in the 2+1d $\mathbb{Z}_2$ lattice gauge theory. To do this, we begin begin with a discussion of the action and Hamiltonian formulations of the $\mathbb{Z}_2$ lattice gauge theory in Sec.~\ref{sec:mod}. This is followed in Sec.~\ref{sec:theory}, with a review of the path integral matching algorithm. Estimates of the required quantum resources are in Sec.~\ref{sec:resources}. Within Sec.~\ref{sec:res} are numerical results obtained on a quantum simulator for $3^2$ and $4^2$ lattices, and we conclude and consider future work in Sec.~\ref{sec:con}.


\section{Model}
\label{sec:mod}
\begin{figure}[!t]
    \centering
    \includegraphics[width=0.48\textwidth]{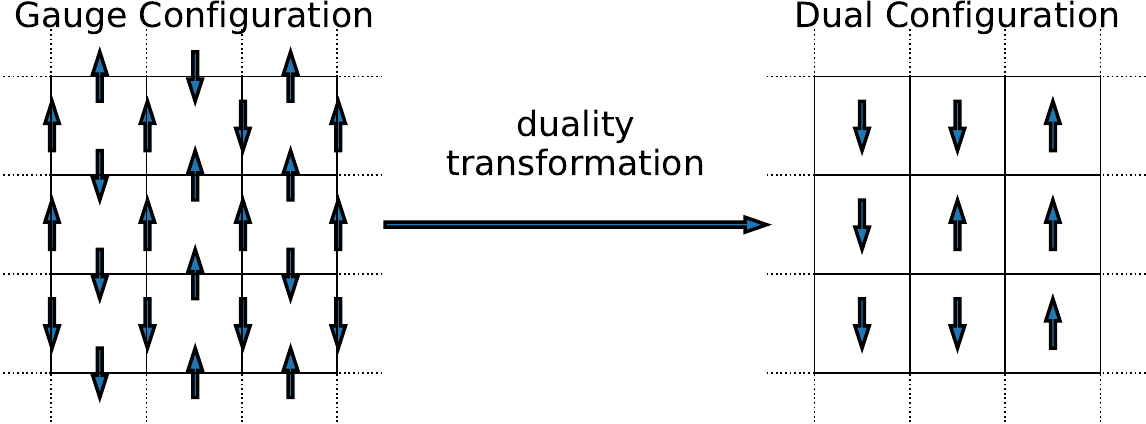}
    \caption{Graphical depiction of the duality transformation on a portion of a generic time slice of the $Z_2$ gauge configuration to a generic $Z_2$ spin configuration time slice. The up arrow corresponds to a positive spin.}
    \label{fig:dualitytransformation}
\end{figure}

Simulations on quantum computers are naturally formulated in the language of Hamiltonians, while classical, Euclidean lattice field theory uses actions, therefore we must specify both in order to perform calculations. We begin our discussion with the anisotropic Wilson action for a general gauge theory which can be written as:
\begin{equation}
\label{eq:wact}
    S=-\beta_t\sum_{t}\re\Tr U_t-\beta_s\sum_{s}\re\Tr U_t
\end{equation}
where $t,s$ label the temporal and spatial directions and the plaquettes $U_i$ are formed from gauge links given by elements of the group.  After gauge fixing the time-like links, we will classical sample with this action.  Using $\beta_t\neq\beta_s$ introduces an anisotropy $\xi=a/a_0$ where the physical lattice spacing in the temporal direction $a_0$ is not equal to the spatial one $a$. From Eq.~(\ref{eq:wact}), we can derive the Kogut-Susskind Hamiltonian~\cite{PhysRevD.11.395} via the transfer matrix in the limit of $\xi\rightarrow\infty$  (see~\cite{Creutz:1984mg,Lamm:2019bik}).  This Hamiltonian is given by:
\begin{equation}
\label{eq:latham}
    H=-c\left[\frac{1}{\beta_H}\sum_{\{ij\}}l_{ij}^2-\beta_H\sum_{s}\re\Tr U_s\right]
\end{equation}
where $l_{ij}$ are the conjugate variables of $U_s$. We have introduced the Hamiltonian coupling $\beta_H=\sqrt{\beta_s\beta_t}$ and the bare speed of light $c=\sqrt{\frac{\beta_s}{\beta_t}}$.  While efficient digitization of gauge groups is a field of active research~\cite{Zohar:2012ay,Zohar:2012xf,Tagliacozzo:2012df,Zohar:2013zla,Zohar:2015hwa,Zohar:2016iic,Hackett:2018cel,Bazavov:2015kka,Bender:2018rdp,Zhang:2018ufj,Unmuth-Yockey:2018xak,Unmuth-Yockey:2018ugm,Zache:2018jbt,Raychowdhury:2018osk,Kaplan:2018vnj,Zohar:2018nvl,Bender:2018rdp,Luo:2019vmi,Alexandru:2019nsa,Klco:2019evd,Ji:2020kjk,Brower:2020huh,Kreshchuk:2020dla,Kreshchuk:2020kcz,Haase:2020kaj,Davoudi:2020yln}, specializing to the case of $\mathbb{Z}_2$ gauge theory is straight forward because we map each link to a single qubit. It then follows that Eq.~(\ref{eq:latham}) becomes:
\begin{equation}
    \label{eq:GaugeHamiltonian}
    \hat{H}_{\text{gauge}} = -c\left[\frac{1}{\beta_H} \sum_{i \in \text{links}} \hat{\sigma}^x_i- \beta_H \sum_{s} \prod_{i \in s} (\hat{\sigma}^z_i)^{\otimes}\right],
\end{equation}
Although this mapping is relatively simple, the plaquette-term requires a four-qubit operation.  This can be avoided by reformulating this theory in its dual representation -- the transverse Ising model in 2+1d~\cite{Wegner:1984qt,Kogut:1979wt,Yamamoto:2020eqi}. The relation between the theories maps the flux on the plaquette to a spin on the dual lattice \cite{Wegner:1984qt,Kogut:1979wt,Yamamoto:2020eqi} and is graphically depicted in \Fref{fig:dualitytransformation}.  

In this representation, the plaquette-term becomes a single-qubit operation, while the $l_{ij}$ term becomes a two-qubit one:
\begin{equation}
    \label{eq:DualHamiltonian}
    \hat{H}_{\text{dual}} = - J \sum_{\vec{n}} \hat{\sigma}^x_{\vec{n}}\hat{\sigma}^x_{\vec{n}+\hat{\mu}}- \Gamma \sum_{\vec{n}} \hat{\sigma}^z_{\vec{n}}\equiv H_K+H_V,
\end{equation}
where the summations over $\vec{n}$ correspond to the centers of the plaquettes in the gauge representation and the sum over $\hat{\mu}$ is the unit vectors in the $\hat{x}$ and $\hat{y}$ directions. The relation between the two representations is seen in~\Fref{fig:dualitytransformation}. The couplings are related by $\Gamma=c\beta_H$ and $J=\frac{c}{\beta_H}$. This dual representation with periodic boundary conditions effectively halves the number of qubits because all degrees of freedom are gauge-invariant.

This gauge theory has a confined and deconfined phase. When $J \ll \Gamma$ ($\beta_H\gg 1$), the model is deconfined and excitations correspond approximately to plaquettes with flux pointing in the opposite direction. On the other hand when $\Gamma \ll J$ ($\beta_H\gg 1$), the system is confined and excitations correspond to droplets with domain walls in the $\hat{\sigma}^x$ basis. In this work, we use one set of couplings: $J = 0.3$ and $\Gamma = 1$.

The final object we need to define for our model is the projection operators for particle-like states. For this paper, we will investigate quantum states of fixed parity excited by the operator
\begin{equation}
    \label{eq:plainwaveexcitation}
    a^{\dagger}_{\textbf{k}} = \sum_{\textbf{r}} f(r) e^{-i\, \textbf{k}\cdot\textbf{r}} \hat{\sigma}^+_{\textbf{r}}
\end{equation}
where $\textbf{k}=2\pi[\frac{k_x}{n_x},\frac{k_y}{n_y}]$, $\textbf{r}=[r_x,r_y]$.  Due to the small lattices, we only consider plane wave excitations i.e. $f(r)=1$.  As larger lattices become available, it would be interesting to study how wave packets with nontrivial envelopes $f(r)$ evolve since these should have superior overlap with physical particles.



\section{Algorithm}
\label{sec:theory}
We are interested in the matrix elements $\langle \psi_i |\mathcal{O}(t)|\psi_j\rangle$ of operators $\mathcal{O}(t)=e^{iHt}\mathcal{O}e^{-iHt}$ between two states, $\psi_i$ and $\psi_j$. The difficulty in preparing these strongly-coupled states on the quantum computer can be avoided by instead computing the matrix elements $\langle \Psi_i |O(t)|\Psi_j\rangle$ between basis states $\Psi_i$ which are cheaper to prepare, and then weighting various matrix elements properly.  This forms the basis of our hybrid algorithm-- the classical portion is used to obtain the weights, and the quantum potion computes matrix elements between easily prepared basis states. 

To do this, we consider a thermal state given by a density matrix $\rho\equiv e^{-\beta \hat{H}}=\sum_{ij}\rho_{ij}|\Psi_i\rangle\langle\Psi_j|$ with a Hamiltonian $H$, inverse temperature $\beta$, and $\rho_{ij}=\left<\Psi_i\right|\rho\left|\Psi_j\right>$.  Provided this state has overlap with $\psi_i,\psi_j$, then there are two operators $P,Q$ respectively that would project out the component of the thermal state.  Thus, if one can properly sample from $\rho$, then the desired matrix elements can be written as
\begin{align}\label{eq:expectation}
\left<\psi_i|\mathcal O(t)|\psi_j\right> &= \frac{\Tr P^{\dag} e^{-\beta H} Q \mathcal O(t)}{\Tr e^{-\beta H}}\notag\\
&=\frac{\sum_{i,j}(P^{\dag}\rho \,Q)_{ij}\mathcal O(t)_{ji}}{\sum_i\rho_{ii}}
\equiv\frac{\langle Q\mathcal O(t) P^{\dag}\rangle_{\rho}}{\langle \delta_{ij}\rangle_{\rho}}
\end{align}
where  $\mathcal O(t)_{ji}=\left<\Psi_j\right|\mathcal O(t)\left|\Psi_i\right>$. The notation $\langle \mathcal \cdot\rangle_\rho$ denotes expectation values sampled from the distribution $\rho_{ij}$.  The overall normalization $\langle \delta_{ij}\rangle_{\rho}$ measures the weight of $\sum_i\rho_{ii}$ which is often unneeded, but can be computed if desired~\cite{Harmalkar:2020mpd}.

Efficient classical sampling of the distribution $\rho_{ij}$ can be obtained from standard Euclidean path integral methods provided open boundary conditions (OBC) in time are used~\cite{Luscher:2011kk}. The classical side of the algorithm thus involves Monte Carlo simulations to sample the $\rho$ in the gauge representation. The configurations are then transformed to the dual representation and using \Fref{eq:plainwaveexcitation} for $P,Q$ we obtain the initial and final states. These operators $P,Q$ naturally belong to the classical portion of the algorithm because they are non-unitary projections. However it is possible to include them in the quantum portion using projective measurements such as \cite{Barath}. 

The quantum portion of the algorithm implements a measurement of $\langle\Psi_i| \mathcal{O}(t)|\Psi_j \rangle$. Matrix elements of the form $\langle\Psi_i| \mathcal{O}(t)|\Psi_i \rangle$ may be efficiently computed on a quantum processor~\cite{PhysRevLett.118.010501,PhysRevLett.123.070503,Roggero:2018hrn,Zohar:2018cwb,Clemente:2020lpr}; thus we recast our matrix elements in terms of diagonal ones $|\Psi_u\rangle,|\Psi_v\rangle=|\Psi_i\rangle\pm|\Psi_j\rangle$.  If these are instead used as the initial states on the quantum computer, the desired matrix elements can be obtained via
\begin{equation}
    \langle \Psi_i| \hat{\mathcal{O}} |\Psi_j\rangle + \langle \Psi_j| \hat{\mathcal{O}} |\Psi_i\rangle =\langle \Psi_u| \hat{\mathcal{O}} |\Psi_u\rangle-\langle \Psi_v| \hat{\mathcal{O}} |\Psi_v\rangle.
\end{equation}

\section{Quantum Resources}
\label{sec:resources}
Key to the effectiveness of this method is the cost of preparing $|\Psi_u\rangle,|\Psi_v\rangle$ on given quantum hardware is reduced compared to the full $|\psi_i\rangle,|\psi_j\rangle$. To study this, we use CNOTs as the universal two-qubit gate and compute how many are required to implement these states. 

We consider two ways of preparing the initial superposition states $|\Psi_u\rangle$ and $|\Psi_v\rangle$ which each have different costs and benefits. One method which reduces the quantum cost is to perform a quantum simulation for each summand $f(r)e^{-i\textbf{k}\cdot\textbf{r}}\sigma^+_\textbf{r}$ of \Fref{eq:plainwaveexcitation} and then classically perform the sum over $\textbf{r}$. The benefit of this is the cost of preparing the state is $O(N_s^2)$ two-qubit gates where $N_s$ is the lattice size in one dimension.  A downside to this method is the number of circuits needed is $(N_s^4)^{N_{p}}$ where $N_p$ is the number of excitation operators $a_\textbf{k}^\dag$ used. While this is tractable for few particles, it clearly scales poorly asymptotically. While this method is $O(N_s^2)$, the overall coefficient is always less than one. This is because in the case of $\mathbb{Z}_2$, the dual representation has only two spin states per site. Thus, $|\psi_i\rangle$ and $|\psi_j\rangle$ can differ by at most $N_s^2 / 2$ sites, depending on correlations between sites. Only in the case of differences are two-qubit operations required; therefore, the number of CNOT gates is $\approx c(J,\Gamma)N_s^2/2$ where $0\leq c(J,\Gamma)\leq1$ is a coupling-dependent number related to the fraction of sites that actually differ.

Averaging over the different $\textbf{k}$ which are allowed, the number of CNOT gates required for the set of couplings considered for small lattices are listed in Table \ref{tab:statecosts}. From these results, we have confirmation of the $O(N_s^2)$ from a direct implementation for the split summation method. Using this data, we can estimate the asymptotic $c(0.3,1)\approx0.7$.

\begin{table}[ht]

 \caption{The average number of CNOT gates for preparing one ($1p$) and 2 ($2p)$ plane waves on $N_s^2$ lattice. 100 configurations were used in total to extract this estimate.}
    \label{tab:statecosts}
\begin{center}
    \begin{tabular}{ c c c}
    \hline\hline
         $N_s^2$ & $1p$ CNOTs & $2p$ CNOTs \\ \hline 
         $2^2$ & $2 \pm 1$ & $2 \pm 1$ \\ 
         $3^2$ & $2\pm1$ & $3\pm1$ \\
         $4^2$ & $3\pm1$ & $4\pm1$ \\
         $5^2$ & $4\pm1$ & $5\pm1$ \\
         $6^2$ & $7\pm1$ & $8\pm1$ \\
         $7^2$ & $10\pm1$ & $11\pm1$\\
         \hline \hline
    \end{tabular}
    \end{center}
\end{table}

The second method uses an ancilla qubit. Suppose we have a pair of unitary operations, $\hat{U}_i$ and $\hat{U}_j$, that prepare the initial states $|\psi_i\rangle$ and $|\psi_j\rangle$. These unitaries are expected to scale like $O(N_s^{2 N_{p}})$ if we want to prepare a complete plane wave but it is possible this would be $\mathcal{O}(N_s^2 \log(N_s)+N_p)$ if a Fourier transform can be used \cite{PhysRevLett.113.010401,Kivlichan_2020}. If the ancilla is prepared in the $|\pm\rangle = |0\rangle \pm |1\rangle$ state then controlled implementations of the unitaries can be applied and then transforming the ancilla back to the computational basis will efficiently add or subtract the two states. When the spatial dimensions have an even number of plaquettes, we can write the trotterization of the time evolution operator $U(t)$ up to $O\Big(N_t \delta t^2\Big)$ as a stroboscopic set of operators, alternating between even and odd site spins  \cite{Lloyd1073}:
 \begin{equation}
    \label{eq:trotterization}
    \begin{split}
    \hat{U}(t) &= e^{-i t \hat{H}_{dual}} \approx \left[\hat{U}_V(\delta t)\prod_{N,i}\hat{U}_{N,i}(\delta t)\right]^{N_t},
    \end{split}
\end{equation}
$U_V$ corresponds to the potential energy $H_V$ in Eq.~(\ref{eq:DualHamiltonian}) and $U_{N,i}$ correspond to kinetic energy in the $\textbf{n}=\textbf{x},\textbf{y}$ direction for $i=e,o$ even or odd sites:
\begin{equation}
\label{eq:operators}
    \begin{split}
    U_V(\delta t) & = e^{-i \delta t \sum_{\vec{n}}\hat{\sigma}^z_{\vec{n}}}\\
    U_{N,i}(\delta t) & = e^{-i \delta t \sum_{\textbf{n},~i}\hat{\sigma}^x_{\textbf{n}}\hat{\sigma}^x_{\textbf{n} + \textbf{x}}}\\
    \end{split}
 \end{equation}
 where $e,o$ indicate the even or odd spatial sites used to decompose the lattice. $\hat{U}_V(\delta t)$ is a product of single qubit rotations, and thus may be applied in one step. The operators $\hat{U}_{N,i}(\delta t)$ required two-qubit gates and thus depend on gates available. Together, the four $U_{N,i}$ require $2 N_{\text{s}}^2$ two-qubit gates per Trotter-step if a CNOT or CZ gate are native.  A reduction in cost to $N_s^2$ can be obtained if the Moeller-Sorenson $XX$ is available. While the gate count is roughly fixed, the circuit depth is dependent upon the observable investigated. For Hermitian observables (e.g. magnetization) then the two-qubit operations can be parallelized and the circuit depth will be approximately $4-10$ two-qubit gates deep depending on the native quantum gates and spatial
 dimensions of the lattice. If a unitary such as $\langle \Psi_i | U(t) |\Psi_j\rangle$ is desired, then the two-qubit operations cannot be parallelized and $\sim N_{\text{qubits}}$ Toffoli gates will be required instead because controlled time evolution operators will be necessary to measure this operator. These controlled evolution operators arise from needing to use an ancillary to measure the expectation value of this unitary non-Hermitian operator. Since $\langle \Psi_u|\hat{U}(t)|\Psi_u\rangle$ is not Hermitian, it is not directly accessible from a quantum computer. However if we prepare our system in the state
 \begin{equation*}
     |\Psi\rangle = \frac{1}{\sqrt{2}}\Big(|0\rangle_a |\psi\rangle + |1\rangle_a |\psi\rangle\Big),
 \end{equation*}
 and then apply a controlled version of $\hat{U}(t)$, $CU(t)$ and then measure either $\hat{\sigma}^x$ or $\hat{\sigma}^y$ on the ancilla qubit we will get the real and imaginary parts respectively \cite{Lamm:2019bik}.
 Turning our regular evolution operator into a controlled evolution operator simply involves two transformaitons. The first makes all single qubit unitaries into controlled unitaries, which can easily be done with two CNOT gates, and a few single qubit rotations. Turning the two qubit operations into controlled unitaries simply involves turning all CNOT gates into Toffoli gates and all single qubit operations into controlled ones.
As a simple metric for comparing the path integral matching procedure, we use the cost of a Trotterization step as a proxy for the gate cost of adiabatic state preparation. Clearly, the $c(J,\gamma)N_s^2/2$ two-qubit gate cost of preparing the one- or two-particle states is cheaper than the $2N^2_s$ required single Trotter step. Since adiabatic state preparation typically requires multiple Trotter steps, this suggests for these states, the path integral matching algorithm yields shallower circuits, albeit at the cost of increased classical resources and the total number of quantum circuits.  



\section{Results}
\label{sec:res}
As a demonstration of how particle states can be studied with the path integral matching algorithm we study the one- and two-particle plane wave excitations. In the simulations we examine the system in the deconfined phase with 100 configurations generated with coupling parameters $J = 0.3$ and $\Gamma = 1$. Additonal information regarding the various choices of quantum simulation are listed in Table \ref{tab:timefacts}.
\begin{table}[ht]
    \caption{Information for the quantum simulation of the one- and two-particle state evolution: the parity of the state, lattice size $N_s$, trotterization step $\delta t$, total number of steps $N_t$.}
    \begin{center}
    \begin{tabular}{c@{\hskip 0.2in}c@{\hskip 0.2in}c@{\hskip 0.2in}c}
        \hline\hline
         Parity & $N_s$ & $\delta t$ & $N_t$\\ \hline
         odd &  $3,4$ & 0.15 & 800\\
         odd &  $3,4$ & 0.20 & 600\\
         odd &  $3,4$ & 0.25 & 500\\
         odd &  $3,4$ & 0.3 & 400\\ 
         odd &  $3,4$ & 0.4 & 200\\ 
         \hline
         even & $4$ & 0.05 & 1200\\
         \hline\hline
    \end{tabular}
    \end{center}
    \label{tab:timefacts}
\end{table}

Different choices of states $\psi_i$ (e.g. a pion, two protons) and operators $\mathcal{O}(t)$ (e.g. electromagnetic or axial currents $J_\mu$) the matrix element $\langle \psi_i|\mathcal{O}(t)|\psi_j \rangle$ will correspond to transition form factors and cross-sections. Here, we consider the unitary operator 
\begin{equation}
\label{eq:unitaryoperator}
    \mathcal{O}(t) \equiv U(t)^\dag U(t) U(t) = e^{-i t \hat{H}}.
\end{equation}
In order to properly normalize the energies, we must subtract the energy of the vacuum state $E_V$.  This is done by modifying Eq.~(\ref{eq:unitaryoperator}) to
\begin{equation}
\label{eq:uopvac}
    \mathcal{O}(t) \equiv e^{-i t \hat{H}}e^{i t E_V}=  e^{-i t (\hat{H}-E_V)}.
\end{equation}
where $N_s$-dependent $E_V$ can accurately be measured using classical methods. In this work, we will focus on the extracting the one- and two-particle energies. For this, our projection operators $P=Q$ and for the case of exact time evolution, the matrix elements for Eq.~(\ref{eq:unitaryoperator}) are
 \begin{align}
 \label{eq:mexvac}
     \langle \psi_i|\mathcal{O}(t)|\psi_j \rangle &= \sum_{k,l} \langle \psi_i | \phi_k \rangle\langle \phi_k|O(t)|\phi_l \rangle \langle \phi_l | \psi_j \rangle \notag\\
     &=\sum_{k,l} \langle \psi_i | \phi_k \rangle\langle \phi_k|\phi_l \rangle \langle \phi_l | \psi_j \rangle e^{i t (E_l-E_V)}\notag\\
     &=\sum_{l} \langle \psi_i | \phi_l \rangle \langle \phi_l | \psi_j \rangle e^{i t (E_l-E_V)}\notag\\
     &=\sum_{l} \langle \beta|P | \phi_l \rangle \langle \phi_l | P^\dag|\beta \rangle e^{i t (E_l-E_V)}.
 \end{align}
where $\phi_l$ are the eigenbasis of $\hat{H}$, and $|\beta\rangle$ is the thermal state produced by the stochastic sampling. For the full ensemble, we compute Eq.~(\ref{eq:mexvac}) for multiple values of $t$ and perform a Fourier transformation so that we can extract the spectral function $f(aE)$. Using this spectral function we can extract the energy of given particle states. While the low-momentum particle states could be obtained from Euclidean calculations or exact diagonalization, we use this computation as a nontrivial test of our method to reproduce these results.  Further, extracting these quantities may provide an efficient way toward setting the physical scale of $\delta t$ in Minkowski lattice field theory similar to how the Euclidean $a$ is determined by measuring the Sommer parameter, string tension, or a mass scale.

While Eq.~(\ref{eq:mexvac}) is correct in the limit of $\delta t\rightarrow 0$, Trotterization process introduces additional interactions, which allow for the mixing of states. This can be seen by comparing the generic leading-order Trotter operator to the exact evolution: 
\begin{equation}
    e^{-i(H_A+H_B)t} \approx \left(e^{-iH_A\delta t}e^{-iH_B\delta t}e^{-i[H_A,H_B]\delta t}\right)^{t/\delta t}
\end{equation}
By neglecting the commutator terms like $[H_A,H_B]$, we are effectively simulating a different Hamiltonian which may have reduced symmetries. This means that even in the case where the projection operators exactly pick out a single state, finite $\delta t$ calculations of Eq.~(\ref{eq:mexvac}) may contain contamination from other states -- potentially even ones with the incorrect quantum numbers. 

\subsection{Dispersion Relation: single spin}
\begin{figure}
    \centering
    \includegraphics[width=0.48\textwidth]{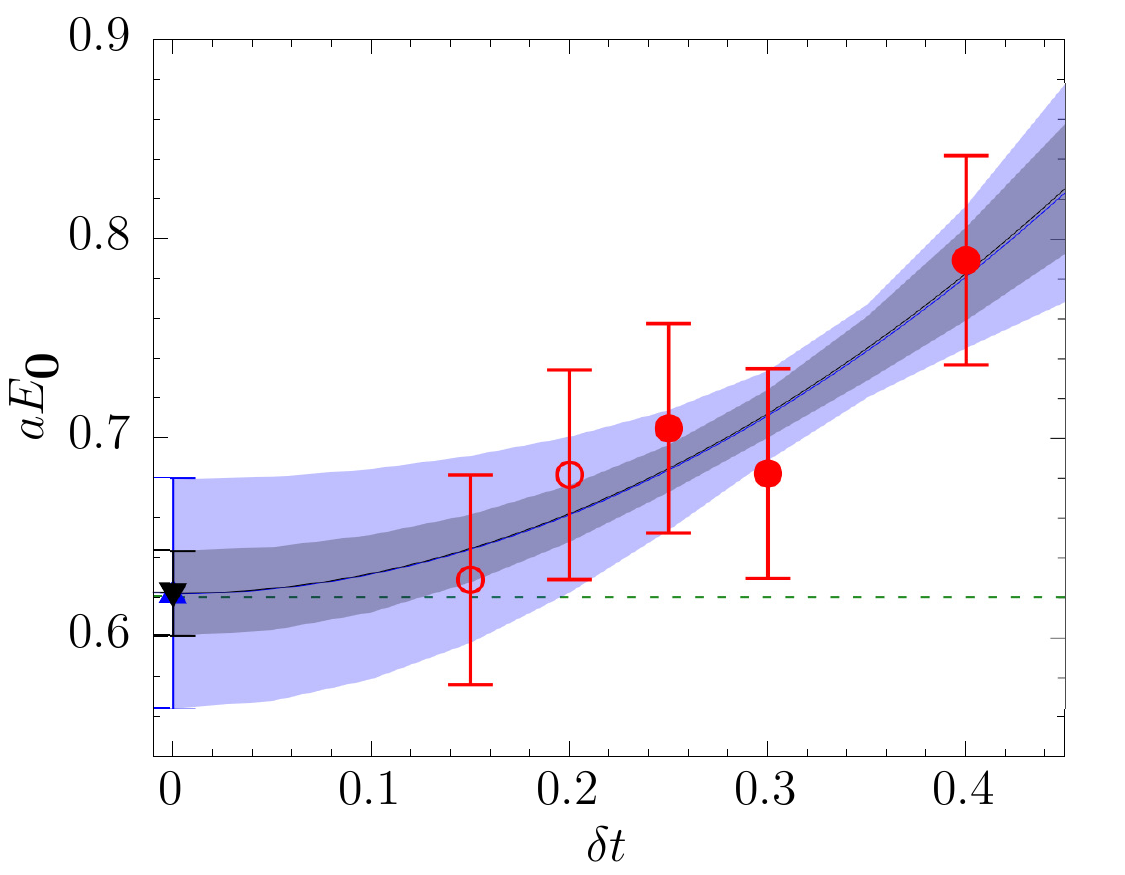}
    \caption{Results for $aE_{\textbf{0}}$ vs $\delta t$ on a $3^2$ lattice.  The $\delta t\rightarrow 0$ values of (\protect\bluetri) and (\protect\blacktri) use only $\delta t\geq0.25$~(\protect\redfcir) or with $0.15\leq\delta t\leq0.25$ (\protect\rbcir) respectively.  The dashed line indicates the $\delta_t\rightarrow0$ result.}
    \label{fig:fitextrap}
\end{figure}

First, we computed the matrix element $ \langle \textbf{k}| \mathcal{O}(t) |\textbf{k}\rangle$ of the states excited by using \Fref{eq:plainwaveexcitation} for fixed momenta $\textbf{k}$ as a projection operator on $N_s^2=3^2,4^2$ lattices. These states were then evolved for an approximately fixed $T=N_t\delta t$ for $\delta t=0.15,0.2,0.25,0.3,0.4$.  For each $\delta t$,  we perform a discrete Fourier transform with fixed $\delta E$ to obtain an approximation of the spectral function.  For each $\delta t$, we take the peak with largest spectral weight to be the single particle energy $aE_{\textbf{k}}(\delta t)$. An example of these results is shown in Fig.~\ref{fig:fitextrap} for the $\textbf{k}=0$ state.  The dominant error is from the finite $\delta E$ below which we cannot resolve the peaks of the spectral function. This error could be reduced by taking a longer $T$.  In order to extract the $\delta t\rightarrow 0$ results, we perform an extrapolation.  Since the error from the Hamiltonian and the Trotterization should be $O(\delta t^2)$, we fit the data to the function
\begin{equation}
    aE_{\textbf{k}}(\delta t)=c_1+c_2(\delta t)^2.
\end{equation}

To investigate the potential for circuit depth reduction by using larger $\delta t$, we consider two fits.  The first is performed only for the points $\delta t\geq 0.25$ with the result of $aE_{\textbf{0}}=0.62(6)$, while the second is performed including all the data and finds $aE_{\textbf{0}}=0.62(3)$.  Both of these results are in good agreement with the exact $3^2$ value of $aE_{\textbf{0}}=0.6204$.  While obviously using the smaller values of $\delta t$ provide for reduced uncertainty, they come at the cost of larger circuits. For the same fixed $T$, we increased our longest circuit depth by a factor of 1.6.

\begin{figure}
    \centering
    \includegraphics[width=0.48\textwidth]{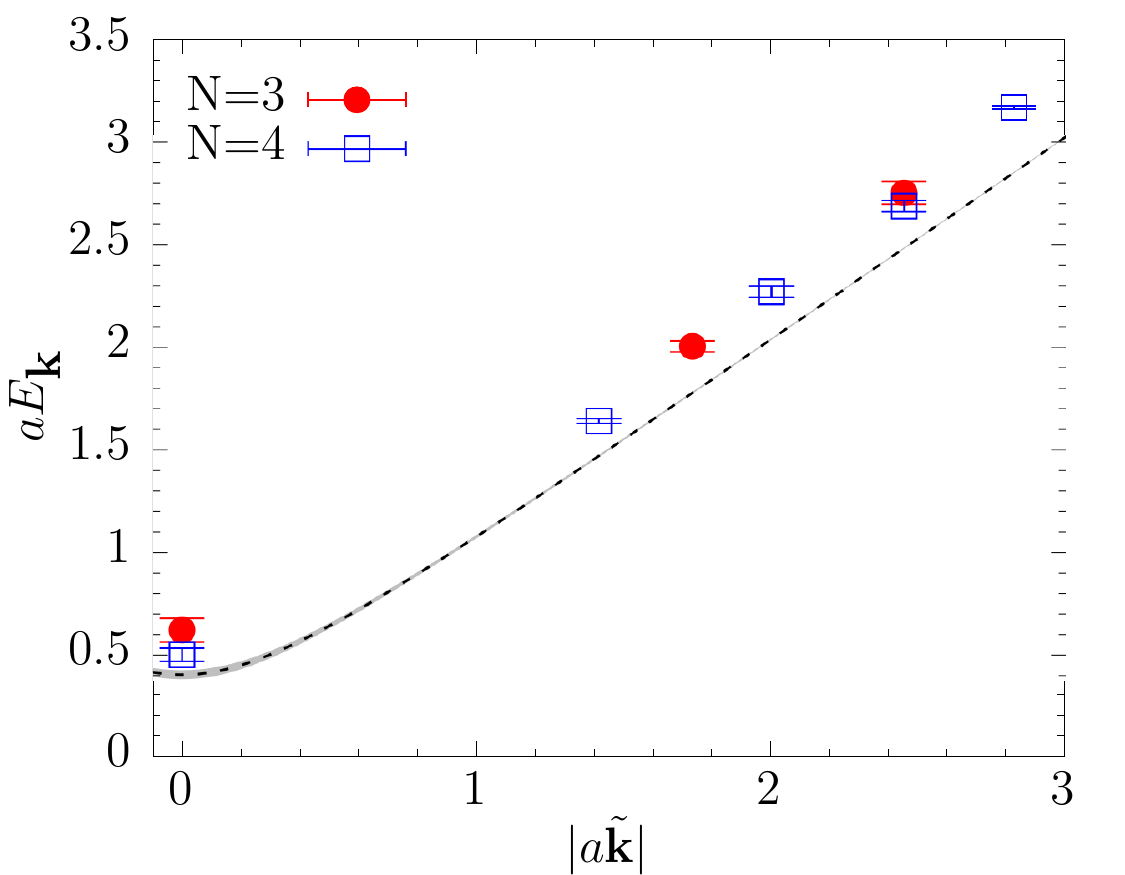}
    \caption{Results for one particle dispersion relation $aE_{\textbf{k}}$ vs $|a\tilde{\textbf{k}}|$ after $\delta t\rightarrow 0$ extrapolate for a spatial lattice of (\protect\redfcir) $3^2$ and  (\protect\bluesq) $4^2$.  The black dashed line corresponds to the continuum dispersion relation $aE_{\textbf{k}}=\sqrt{(am)^2+(a\textbf{k})^2}$.}
    \label{fig:3x3dualsingleparticle}
\end{figure}

In Fig. \ref{fig:3x3dualsingleparticle} we show the dispersion relation obtained by fitting only the $\delta t\geq 0.25$ for all accessible values of the lattice momenta
\begin{equation}
    |a\tilde{\textbf{k}}|\equiv\sqrt{\sum_s\left[2\sin\left(\frac{a\textbf{k}_s}{2}\right)\right]^2}
\end{equation}
where $s$ sums over the spatial directions.
We also plot a continuum dispersion relation $aE_{\textbf{k}}=\sqrt{(am)^2+(a\textbf{k})^2}$ where $am=0.40(2)$ is given by extrapolating the $N_s=2,3,4$ results of $aE_{\textbf{0}}=0.8701,0.62(2),0.50(3)$ assuming $O(a^2)$ corrections dominate, as suggested by \cite{Agostini:1996xy}. The $N_s=2$ value was computed by exact diagonalization for simplicity.  Comparing $am$ to these values, we can see that for $N_s\leq 4$ there are substantial finite volume effects at $k=0$, which increase for $\textbf{k}>0$.  Despite this, we find qualitative agreement between our results for the dispersion relation at finite volume and the continuum extrapolated one. 

\subsection{Two-particle states}
\begin{figure*}
    \centering
    \begin{picture}(500,200)
    \put(0,0){\includegraphics[width=\textwidth]{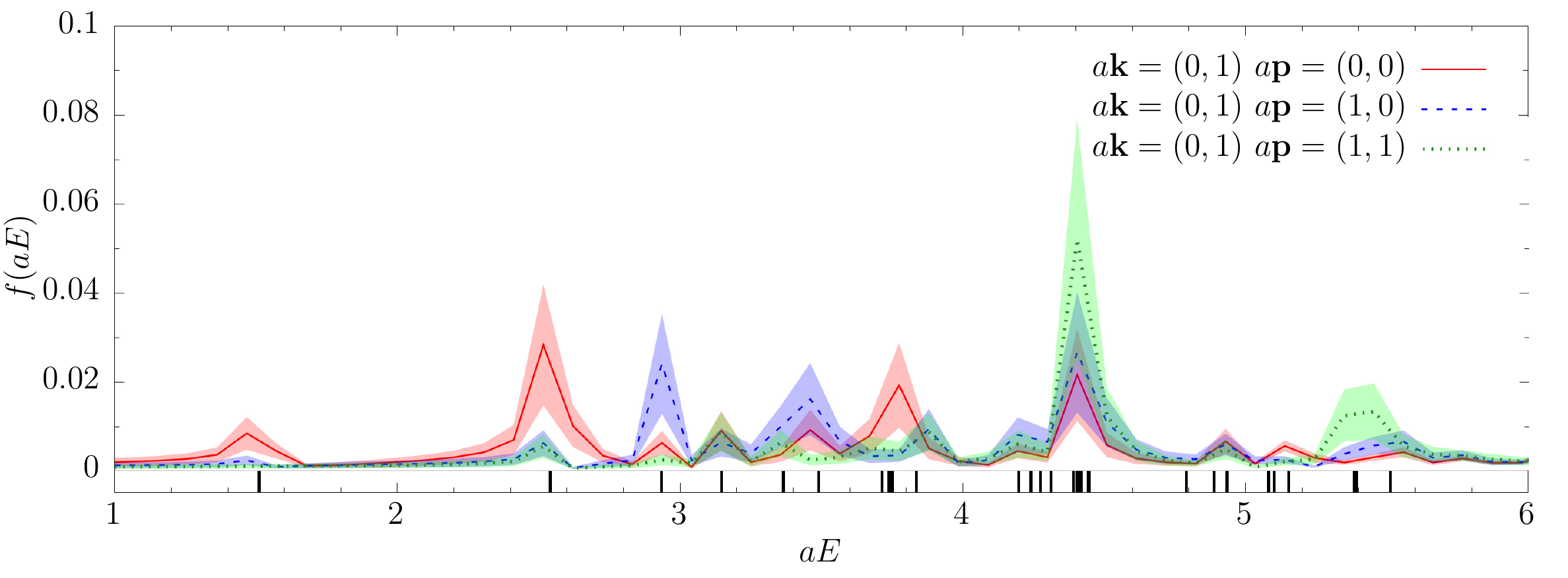}}
    \put(45,84){\includegraphics[width=0.45\textwidth]{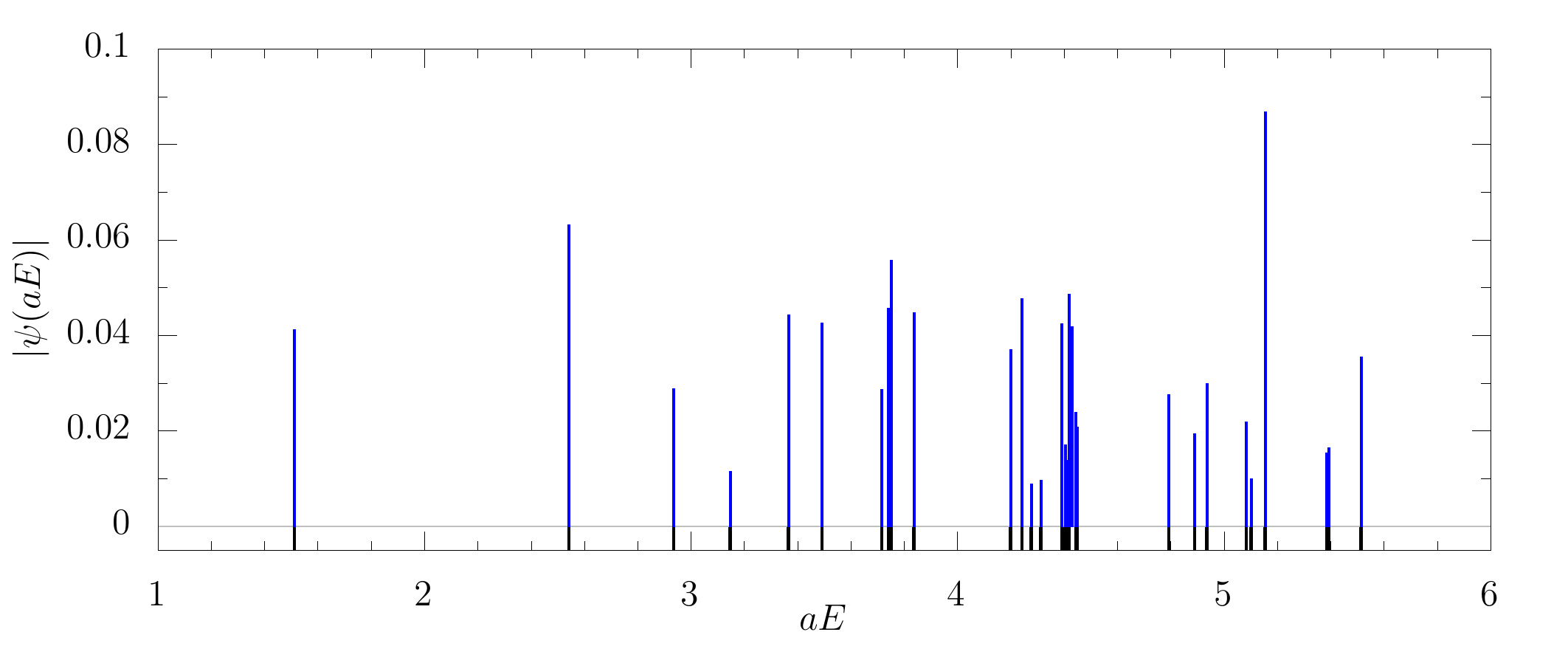}}
    \end{picture}
    \caption{The two-excitation spectral function for various initial momenta on a $4 \times 4$ lattice. The black ticks underneath the plot correspond to the energies found via exact diagonalization. (inset) The eigenstate decomposition of the $a\textbf{k}=(0,1)$, $a\textbf{p}=(1,0)$ initial state}
    \label{fig:4x4dualtwoparticle}
\end{figure*}

Next, we considered the case of two plane-waves scattering by computing the matrix element $\langle\textbf{k},\textbf{p}|O(t)|\textbf{k},\textbf{p}\rangle$ excited by $\hat{P} = \hat{a}^{\dagger}_{\textbf{k}} \hat{a}^{\dagger}_{\textbf{p}}$ for $N_s^2 = 4^2$ and $\delta t=0.05$. Our final results are the spectral function shown in Fig. \ref{fig:4x4dualtwoparticle} obtained from a discrete Fourier transform.

In order to demonstrate time dependence, we need to compare these spectral functions to the initial state of the scattering plane waves.  In the inset of Fig. \ref{fig:4x4dualtwoparticle}, we present the eigenstate decomposition for the initial state of $a\textbf{k}=(0,1)$, $a\textbf{p}=(1,0)$, which we can compare to the final result in the larger figure. We find a change in the relative weight of various eigenstates from $t=0$. This indicates that the wavepackets are interacting and that scattering processes can occur. Alas, we do not observe a clean spectrum at $t=0$ of a single eigenstate. This, in turn, suggests the two plane wave ansatz for the source and sink does not have as strong overlap with two single particle eigenstates on this lattice.  This suggests a need both for larger lattices such that the particle states can be physically separated, and that wavepacket-like excitations should be investigated.

\section{Conclusions}
\label{sec:con}
This work has extended the general methods developed in \cite{Harmalkar:2020mpd} to extract the matrix elements of particle excitations while still reducing the circuit depth compared to adiabatic state preparation.  This has been explicitly shown for the dispersion relation and a two-particle spectral function in the $\mathbb{Z}_2$ gauge theory on small lattices. Similar calculations for this model may be tractable to simulate on quantum computers in the near future, with $N_s=2^2$ lattices potentially feasible already. We have also observed evidence of scattering and particle interactions.  A crucial direction of future work would be to study how other matrix elements, e.g. form factors, can be extracted using these techniques.

While the plane wave ansatzes used here was capable of extracting meaningful results, they are likely not an ideal choice for finite size particle states because their overlap with multiple states will lead to signal to noise problems which, unlike Euclidean calculations, are not suppressed at longer times. More optimal choices for projection operators will be required, as well as signal-to-noise mitigation techniques. In the case of scattering states, improved operators are currently being developed in the Euclidean lattice field theory community~\cite{Bali:2016lva,Egerer:2020hnc} and will be an important avenue of study in the future of the method studied here. Novel techniques in state preparation on quantum devices such as projected cooling~\cite{Lee:2019zze,Gustafson:2020vqg} may also prove useful in reducing excited state contamination.


\begin{acknowledgments}
The authors would like to thank Scott Lawrence, Yannick Meurice, and Yukari Yamauchi for helpful comments on this work.
E.G. is supported by a Department of Energy Grant under Award Number DE-SC0019139. H.L. is supported by a Department of Energy QuantiSED grant. Fermilab is operated by Fermi Research Alliance, LLC under contract number DE-AC02-07CH11359 with the United States Department of Energy.
\end{acknowledgments}

\bibliography{wise}
\end{document}